\documentclass[twocolumn]{article}

\usepackage[utf8]{inputenc} 
\usepackage[T1]{fontenc}   
\usepackage{hyperref}       
\usepackage{url}            
\usepackage{booktabs}       
\usepackage{amsfonts}      
\usepackage{nicefrac}       
\usepackage{microtype}      
\usepackage{graphicx}
\usepackage{amsmath}
\usepackage{subcaption}

\usepackage{tabularx}
\usepackage{upgreek}
\usepackage{chemmacros}

\usepackage{siunitx}
\usepackage{color}

\usepackage{algorithm}
\usepackage{algorithmicx}
\usepackage[noend]{algpseudocode}

\usepackage{doi}
\usepackage[backend=biber, style=numeric, sorting=none,doi=true]{biblatex}
\begin{document}

\title{Noise-balanced multilevel on-the-fly sparse grid surrogates for coupling  Monte Carlo models into continuum models with application to heterogeneous catalysis}

\author{
Tobias Hülser\thanks{huelser@fhi.mpg.de}~ and Sebastian Matera\thanks{matera@fhi.mpg.de} \\
 Fritz Haber Institute of the Max Planck Society, Berlin, Germany
}

\maketitle

\begin{abstract}
Multiscale simulations utilizing high-fidelity, microscopic Monte Carlo models to provide the nonlinear response for continuum models can easily become computationally intractable. Surrogate models for the high-fidelity Monte Carlo models can overcome this but come with some challenges. One such challenges arise by the sampling noise in the underlying Monte Carlo data, which leads to uncontrolled errors possibly corrupting the surrogate even though it would be highly accurate in the case of noise-free data. Another challenge arises by the ’curse of dimensionality’ when the response depends on many macro-variables. These points are addressed by a novel noise-balanced sparse grids interpolation approach which, in a quasi-optimal fashion, controls the amount of Monte Carlo sampling for each data point. The approach is complemented by a multilevel on-the-fly construction during the multiscale simulation. Besides its efficiency, a particularly appealing feature is the ease of use of the approach with only a single hyperparameter controlling the whole surrogate construction - from the surrogate’s accuracy with guaranteed convergence to which data needs to be created with which accuracy. The approach is demonstrated on challenging examples from heterogeneous catalysis, coupling microscopic kinetic Monte Carlo models into macroscopic reactor simulations.
\end{abstract}
\textbf{\\Keywords:} Heterogenous Multiscale Modeling, Stochastic Simulators, Computational Fluid Dynamics, Microkinetic Modeling, Error control
\newpage
\section{Introduction}
Multiscale phenomena appear in many processes in nature and industry, requiring dedicated modeling and simulation approaches. One class of such approaches is heterogeneous multiscale modeling, where microscopic high fidelity models (HFMs) are employed to close the equations of a coarse-grained model \cite{engquist2007heterogeneous}.
Often this closure goes beyond the determination of some parameters, and the HFM instead serves as a function representing the non-linear system's response to changes in the coarse-grained variables. Corresponding coupled simulations can become computationally intense or even intractable, as HFMs are typically rather costly and need to be evaluated repetitively during the coarse-grained simulation.

Therefore, recent years have seen an increased usage of surrogate models replacing the expensive high fidelity model with a cheaper one, that approximates the true functions on basis of a feasible number of high-fidelity evaluations.
Since the high-fidelity response can be very general, flexible surrogate strategies with tunable accuracy are preferable and a number of different approaches have successfully been applied in the multiscale modeling context, e.g., splines \cite{votsmeier2010simulation}, Shepard interpolation \cite{matera2009first,matera2014predictive,lorenzi2017local}, kernel methods \cite{wirtz2015surrogate,leiter2018accelerated},
random forests \cite{partopour2018random,bracconi2020training}, neural networks \cite{white2019multiscale,doppel2023efficient}
or in situ adaptive tabulation (ISAT) \cite{bracconi2017situ,varshney2005multiscale,varshney2008reduced}.

However, surrogate modeling introduces new challenges, particularly in the context of multiscale modeling. The first major challenge concerns efficiency with respect to the cost of creating the training data, but also - to a lesser extent - with respect to the time a query is answered by the surrogate during the coupled simulation. In this context, an important point is that surrogates often need to approximate functions over high-dimensional domains. This happens, for instance, in  reactive flows with many chemical species, when a HFM delivers the reaction rates as a function of the species concentrations \cite{hulser2025multilevel}. 
In such cases, the surrogate needs to overcome the so-called curse of dimensionality, i.e., the  number of needed ``training'' data should not explode when the dimensionality becomes high \cite{bellman1957dynamic}.
The second major challenge results from the complexity of multiscale modeling, already requiring a high degree of expert knowledge for the involved coarse-grained and HFMs and corresponding simulation approaches. This makes easy-to-use surrogate approaches preferable, which work in an automatic fashion. Ideally, such approach should possess only a single hyperparameter which controls the whole constructions of the surrogate including the ``training set'' design as well as the error of the surrogate. Classical numerical approaches, such as tensor-grid splines,
often come with such strategies, but suffer from the curse of dimensionality. On the other hand, machine-learning approaches often perform well also in higher-dimensional settings but typically come with a large number of hyperparameters with unknown influence on the surrogates accuracy and no clear-cut strategy for ``training set'' design and for improving the accuracy.
Finally, most HFMs require the evaluation by some numerical method and thus the data will carry some error, which, however, can be controlled by the amount of computational effort per data point. The accuracy of a surrogate is thus influenced by two sources of errors: i) the surrogates' discretization error due to the use of a finite number of data points, and ii) the model evaluation error due to the finite amount of CPU time spend to approximate the true HFM response. The latter can completely spoil the accuracy of the surrogate, even though we have enough data points \cite{james2021introduction}. On the other hand, an only coarse surrogate based on rather few data points will not profit from highly accurate and therefore costly  evaluation of the HFM. The accuracy might then be limited as the computational budget is already spent on unnecessarily accurate high-fidelity simulations instead of increasing the amount of data. This is a particular issue when the HFM requires some kind of Monte Carlo sampling, which noise error converges only slowly. An often encountered multiscale setting where this appears is a continuum model which equations are closed by some kind of molecular simulation model, e.g., molecular dynamics \cite{rouet2014spatial,smith2024multiscale,delle2020particle}.
There exist an extensive literature on quantifying the impact of noisy data on the surrogate's accuracy and how to minimize this impact, e.g., \cite{ankenman2010stochastic,le2015asymptotic,wang2019controlling}. 
However, we are aware of only relatively few studies which targeting at balancing sampling with discretization error in surrogate modeling by controlling the sampling effort per data point \cite{dopking2021error,heinrich2001multilevel,chen2017sequential}. 

In this study, we present a surrogate modeling approach which addresses these challenges for Monte-Carlo HFMs. This is based  on our recently developed multilevel on-the-fly sparse grid approach (ML-OTF-SG) originally designed for quasi-exact HFMs \cite{hulser2025multilevel}. This approach utilizes the proven fast convergence of sparse grids (SG) also in high-dimensional settings \cite{bun04,klimke2005algorithm,jakeman2012},
combined with a multilevel on-the-fly construction. This strategy exploits that, during a coupled simulation, typically only a small subdomain  of the space of the coarse-grained variables are visited and highly accurate surrogates are only needed close to the solution in an iterative solution process. Besides achieving high data efficiency, in terms of the number of data points, the ML-OTF-SG strategy possesses only a single hyperparameter, the maximum SG level $L$, controlling accuracy and which high-fidelity data needs to be added. Further, the on-the-fly construction is self-consistent, i.e., an identical query at different stages of the iterative solution process will result in the same response. This is difficult to achieve with general surrogate strategies in conjunction with on-the-fly construction, but also important to ensure numerical stability of the coupled simulation \cite{tynes2025will}. 
We complement this with a balancing of the effects of SG discretization and noisy Monte Carlo simulation results. For this, we decompose the expected squared 2-norm error of the surrogate into a SG discretization error and a sampling error. Balancing both errors allows then to determine the amount of Monte Carlo sampling for each data point. This requires estimates of the discretization error, and we propose a heuristic strategy to obtain these which guarantees convergence. This noise-balancing is done such that the  ML-OTF-SG strategy maintains its convenient properties, with the maximum level $L$ controlling the whole process. Also, the self-consistent on-the-fly construction is maintained, and the solver can be employed in a black-box fashion with almost arbitrary coarse-grained non-linear solvers with guaranteed convergence and a built-in convergence test.

We outline the methodology in Sec. \ref{sec:methods} for an abstract continuum model with a general stochastic model as basis for the HFM. While we expect that the methodology can be useful in more general settings, the continuum setting is particularly suited for an on-the-fly strategy as the coarse-grained variables are themselves functions of space and time. The visited subdomain of the HFM's input space will thus be low-dimensional, in contrast to the input space itself, which can be high-dimensional. In Sec. \ref{sec:multilevel-on-the-fly-sparse-grids}, we review the ML-OTF-SG methodology and, in Sec. \ref{sec:mc-models}, we discuss how to balance discretization and noise error for Monte-Carlo-type models. We demonstrate the methodology on realistic problems from heterogeneous catalysis, where we couple so-called first-principles kinetic Monte Carlo (1p-kMC) models with a stationary fixed-bed reactor model. This is, on the one hand, motivated by the utmost importance of multiscale modeling for heterogeneous catalysis, as chemical understanding typically requires atomic and electronic scale insight, whereas the relevant objectives are macroscopic conversion and selectivity \cite{pineda2022kinetic}. 
Moreover, the very highly non-linear response of 1p-kMC models is very challenging to approximate \cite{matera2019progress}. 
In all our test cases (Sec. \ref{sec:results}), the noise-balanced ML-OTF-SG achieved highly accurate results at modest computational costs.
In Sec. \ref{sec:Discussion}, we discuss our findings and possible future directions of research on improving and testing the methodology.

\section{Methodology} \label{sec:methods}

We consider an abstract continuum problem for an unknown field $F: \Theta  \rightarrow \mathbb{R}^s$, where $\Theta$ is a $Q-$dimensional domain of spatio-temporal coordinates, i.e., $Q\leq 4$, and $s$ is the target field dimensionality, e.g., the number of species in a reactive flow problem. The general equations write
\begin{align}
 \Gamma(F,G)(q)=0 \label{eq:most-general}  \\
 G(q)=g( X(q) ) \label{eq:constitutive_closure}\\
 X(q)=(F(q), \nabla_q F (q), ...) \label{eq:constitutive_arguments}
\end{align}
where $\Gamma$ is a general nonlinear (differential) operator acting on the functions $F(q),G(q)$, which encode the bulk equations as well as the boundary conditions and $q$ are the spatial and temporal coordinates. The spatial gradient is denoted by $\nabla_q$.
The function $G: \Theta \rightarrow \mathbb{R}^{s'}$ is implicitly defined by the field $F$ and a constitutive function (CF) $g$ for which we assumed locality. That is, the value of the field $G$ at a particular position $q$ is a function of the values $X(q)$ given by the field $F$ and (some of) its derivatives. The dots in Equation \eqref{eq:constitutive_closure} imply that $X(q)$ might contain arbitrary spatial derivatives, but, we assume that $X$ has a finite dimensionality $d$. This setting covers a large class of problems ranging from elasticity, where $g$ encodes the relation between stress and deformation gradient, to transport in fluid mixtures, where fluxes depend on the concentrations and their gradients. In section \ref{sec:results}, we investigate stationary convection-diffusion-reaction equations which describe so-called fixed-bed reactors in heterogeneous catalysis, where the CF relates local reaction rates to local concentrations.

Typically, the problem Equation \eqref{eq:most-general} is not analytically solvable, but we have to obtain a numerical approximation via discretization. In most realistic cases, the discretized model is solved using some iterative solver, e.g., some Newton solver. This produces sequences of approximations of the solution $F^{m}_h$ of the field $F$, where $m$ is the iteration index and $h$ indicates the resolution of the discretization, e.g., the grid spacing using uniform grids or a tolerance criterion in adaptive methods. We restrict to iterative schemes which maximally require first derivatives with respect to the fields $F^{m}_h$. An abstract form of such scheme is
\begin{align}
 F^{m+1}_h = \Psi^m_{\Gamma,h}(F^{m}_h,\{{G}_i^m\}_{i=1}^{M_m}, \{{D}^m\}_{i=1}^{M_m} ... ), \label{eq:general-discretised}\\
{G}_i^m=g(\tilde{X}^m_i),~ D^m_i= \nabla_X g(\tilde{X}^m_i),\label{eq:general-discretised-arguments}
\end{align}
where the dots in Equation \eqref{eq:general-discretised} imply a possible dependence on the history of the iterative process and $\nabla_X$ is the derivative with respect to the arguments of the constitutive function. The arguments $\tilde{X}^m_i$ in Equation \eqref{eq:general-discretised-arguments} are obtained from the approximation $F^{m}_h$, e.g., by interpolating to some (auxiliary) discretization node $q_i$. The described abstract scheme covers a variety of discretizations and iterative solvers, including the combination of finite differences and a Quasi-Newton solver employed in section \ref{sec:results}.

A priori, the CF is generally unknown except for some  constraints which can be derived from macroscopic continuum theories \cite{muller2009fundamentals,liu2002continuum}.
A multiscale modeling approach to this problem is to employ a microscopic high-fidelity model (HFM), which parametrically depends on the arguments $X$. We consider the case, when we have a stochastic microscopic model $S$ and the CF is given by the respective conditional expectation subject to a possible nonlinear transform $T$, i.e.,
\begin{equation}\label{eq:constitutive-relation-expectation}
   G(X)=T( E(S|X) ).
\end{equation}
In the examples section \ref{sec:results}, $S$ will results from a Markov jump process describing the interplay of reaction events in the catalytic cycle.

Often, the conditional expectation can not be calculated exactly, but must be estimated by Monte Carlo simulation of the stochastic model. That is, we approximate $G(X)$ by simulating multiple stochastic realizations for fixed parameters $X$ and take the empirical mean. This raises some severe problems in directly coupled computations, where such estimated are used every time when the evaluation of the CF is queried in the iterative scheme \eqref{eq:general-discretised} and \eqref{eq:general-discretised-arguments}. First, sampling approaches, such as molecular dynamics or kinetic Monte Carlo (cf. Sec. \ref{sec:results}) simulations, are computationally rather expensive and converge only slowly. Second, the estimates contain a finite random error which must be controlled to not corrupt the iterative schemes, which commonly are designed for deterministic problems. Since the number of CF evaluations per iteration $M_n$ can be expected to be large, these two points make such direct coupling approaches very impractical. This goes so far that even very simple problems cannot be tackled on today's and foreseeable computer hardware. Furthermore, many nonlinear solvers require derivative information, which often is particularly difficult to obtain from sampling approaches, amplifying the computational load further.

The high computational load due to the CF evaluations can be significantly reduced by the use of surrogate models, i.e., computational models that approximate the true CF and that can be evaluated fast on a computer \cite{ghavamian2019accelerating,wu2020recurrent,biermann2025enabling}.
These usually possess parameters which are adjusted on basis of comparatively few reference evaluations of the CF. These surrogates then replace the true CF in \eqref{eq:general-discretised-arguments} during the continuum simulation.

In the above multiscale setting, a central objective for practical simulations is therefore to find suitable surrogates which require only manageable computational resources for the creation of ``training'' data. A challenge in this respect arises  when the set of input arguments $X$ becomes high-dimensional which tend to lead to excessive amounts of necessary data for classical approaches. To enable smooth workflows, the surrogate should further allow an easy control of its discretization error and of the data creation process, i.e., for which values of $X$ the original CF must be evaluated. Ideally, the surrogate strategy should also be purpose driven, i.e., it should exploit that the surrogate will be used for the queries \eqref{eq:general-discretised-arguments}. To address these points with quasi-exact CF evaluations, we have recently introduced a multilevel on-the-fly sparse grid  approach (ML-OTF-SG) \cite{hulser2025multilevel}, which we will review in Sec. \ref{sec:multilevel-on-the-fly-sparse-grids}. The above described setting
comes with the additional challenge that the high fidelity model only allows for approximate evaluations carrying some stochastic noise. In Sec. \ref{sec:mc-models}, we decompose the total error into a discretization and a noise-induced error for sparse grid interpolation. Balancing both error sources, we extend the ML-OTF-SG with a quasi-optimal control of the sampling effort per data point, leading to a scheme with guaranteed convergence.

\subsection{Multilevel on-the-fly sparse grids}\label{sec:multilevel-on-the-fly-sparse-grids}

Sparse grids (SG) are most popular in the context of partial differential equations and uncertainty quantification \cite{griebel1998adaptive,nobile2008sparse,ma2010,jakeman2012,dopking2018,chassagneux2023learning}.
But sparse grid interpolants (SGI) possess a number of advantages that should make them particularly attractive as surrogates in multiscale modeling. Besides their ability to accurately interpolate also in higher dimensions, this concerns their ease of use, where a single hyperparameter controls the data creation process as well as the discretization error \cite{bun04}. Multilevel on-the-fly sparse grids (ML-OTF-SG) build upon a certain class of SGI for which it is possible to only locally refine the SGI at the query points, i.e., the SGI is constructed on the fly as the queries arrive. Adapting ideas from nonlinear multilevel methods, ML-OTF-SG further exploits that accurate interpolation is not needed far away from the solution \cite{hulser2025multilevel}.

\begin{figure}
    \centering
    \includegraphics[width=0.5\textwidth]{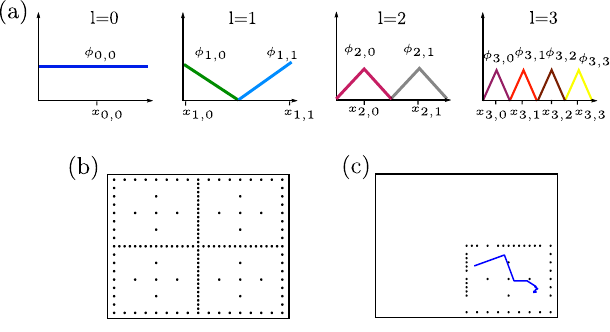}
    \caption{(a) One-dimensional basis functions $\phi_{l,i}$ of different levels $l$ used in the sparse grid construction. (b) Regular sparse grid in 2D. (c) On-the-fly constructed sparse grid, "on-the-fly" evaluation points are depicted as in blue.}
    \label{fig:1D_basis}
\end{figure}

To introduce the ML-OTF-SG methodology, we consider a multidimensional function $f: [-0.5,0.5]^d \mapsto \mathbb{R}^{s'}$. In the above context, this means $f$ is related to the original constitutive function (CF) $g$ by a suitable invertible transform $t$ of its arguments, i.e., $X=t(\mathbf{x})$ and $f(\mathbf{x})=g(t(\mathbf{x}))$. The ML-OTF-SG approach employs a hierarchy of SGIs $f_L$ on the basis of piecewise-linear interpolation \cite{klimke2005algorithm,ma2010,jakeman2012,dopking2018}.
The interpolation level $L$ controls the accuracy of the respective member of the hierarchy, where larger $L$ indicate a more accurate model.
The interpolant $f_L$ is given by
\begin{equation}\label{eq:telescoping-sum}
    f_L(\mathbf{x})=\sum_{\mathbf{l},|\mathbf{l}|_1\leq L} \sum_{\mathbf{i} \in I_\mathbf{l}} \alpha_{\mathbf{i},\mathbf{l}} \phi_{\mathbf{i,l}}(\mathbf{x})
\end{equation}
where $\alpha_{\mathbf{i,l}}$ are the so-called hierarchical surpluses and $\mathbf{l}\in \mathbb{N}_0^d$ (the level index) and $\mathbf{i}\in \mathbb{N}_0^d$ are multi-indices. Each level $\mathbf{l}\in \mathbb{N}_0^d$ corresponds to a set of basis function and a corresponding index set $I_\mathbf{l}$.  The basis functions (BFs) $\phi_\mathbf{l,i}$ are tensor products
\begin{equation}\label{eq:tensor_product_approach}
	\phi_\mathbf{l,i} (\mathbf{x})=\phi_{l_1,i_1}(x_1) \phi_{l_2,i_2}(x_2) ... \phi_{l_d,i_d}(x_d),
\end{equation}
created from 1-dimensional hierarchical piecewise linear finite element BFs
\begin{align}\label{eq:1d-basis-function}
	\phi_{l,i}(x) :=
\begin{cases}
 1,& l=0\\
 \text{max}({1-2^l|x-x_{l,i}|,0}) ,& \text{l>0}
\end{cases}
\end{align}
where $l\geq0$ is the hierarchical level and $i$ distinguishes the different BFs with the same hierarchical level $l$. The 1-dimensional grid points $x_{l,i}$ can recursively be defined by $x_{0,0}=0$, $x_{1,0/1}=\mp 0.5,x_{2,0/1}=\mp 0.25$ and $x_{l,i}=x_{l-1,\lfloor i/2 \rfloor} - (-1)^i/2^l$ for $l>2$. Here, $\lfloor i/2 \rfloor$ is the next smallest integer. The 1-D BFs for the levels $l \in \{0,1,2,3 \}$ are shown in \textbf{Figure} \ref{fig:1D_basis}a.

The SGI $f_L(\mathbf{x})$ is obtained by using all BFs with $|\mathbf {l}|_1 = \sum_i l_i \leq L$ and interpolating the original function $f$ at the corresponding grid points $\mathbf{x}_{\mathbf{l,i}}= (x_{l_1,i_1} \ldots x_{l_d,i_d} )^d$.
That is, we have to evaluate the original function $f$ for all grid points $\mathbf{x}_{\mathbf{l,i}}$ belonging to the d-dimensional BFs $\phi_\mathbf{l,i}$ employed in Equation \eqref{eq:telescoping-sum}. We then obtain the surpluses $\alpha_{\mathbf{i},\mathbf{l}}$ and the SGI $f_L$ by the requirement
\begin{equation}\label{eq:interpolation-requirement}
 f(x_{\mathbf{l,i}})=:f_{\mathbf{l,i}}=f_L(\mathbf{x}_{\mathbf{l,i}}).
\end{equation}
Note that,  SGI restricts the total level of the multidimensional BFs $\phi_{\mathbf{l,i}}$, i.e., $||\mathbf{l}||_1=\sum_i l_i \leq L$. This is in contrast to usual tensor grids allowing all possible products of 1-D BFs. These would arise by dimensionwise restriction of the level, e.g., $l_i \leq L$. As a consequence, a SG will look very different from a classical tensor grid. As an example, we show the grid points in two dimensions for a SG in $L=5$ in Figure \ref{fig:1D_basis}b.
This procedure allows mitigating the ``curse of dimensionality'', i.e., it can be shown that the discretization error $||f -f_{L}||_2$ can be bound by
\begin{align}
	||f -f_L||_2 &\leq C_1 2^{-2L}\cdot L^{d-1},\label{eq:SG_error_bound} \\
	||f -f_L||_2 &\leq C_2 M_L^{-2} \log^{2(d-1)} M_L\label{eq:SG_error_bound2}
\end{align}
for functions with finite mixed second derivatives, where $C_1,C_2$ are constants independent of $L$ and $M_L$ is the number of grid points in the SGI \cite{bun04}. This is in contrast to conventional methods, where the error behaves as $O\left(M^{-r/d}\right)$ (with some constant $r$) and the convergence deteriorates with increasing dimensionality $d$. For a more detailed discussion of SG and particularly the employed SG basis, we refer to the literature \cite{bun04,klimke2005algorithm,ma2010,dopking2018,dopking2022error}.

The SG surrogate $f_L$ then replaces the true CF in the numerical scheme Equation \eqref{eq:general-discretised-arguments} for the solution of the multiscale problem. The ML-OTF-SG approach builds upon these kinds of SGs for that purpose, to reduce the number of evaluations of the expensive function $f$ \cite{hulser2025multilevel}. For this, we first note that the interpolation $f_L(\mathbf{x}_q)$ at a certain query point $\mathbf{x}_q$ is only sensitive to those values $f_{\mathbf{l,i}}$ which BFs $\phi_{\mathbf{l,i}}$ are non-zero at $\mathbf{x}_q$. Since the BFs are only non-zero in an area which shrinks as $2^{-|| \mathbf{l}||}$, this means a large fraction of the data $f_{\mathbf{l,i}}$ is not needed if one is only interested in the interpolation at $x_q$. Now an iterative scheme does a lot of queries to the CF, but these can be expected to lay in only a small subdomain of $[0,1]^d$, i.e., the  possible input values to the high fidelity model. This is particularly obvious when the dimensionality $d$ of the inputs is high. In each iteration, the approximation $F^m_h$ is a function of the $Q\leq 4$ spatial-temporal coordinates. The queries are derived from these functions, and iterative solvers will also create incremental improvements of the approximations. This implies that the queries will visit (the vicinity of) an only low-dimensional domain during the solution process. Combined with the property that evaluation of the SG surrogate on a query point only requires a subset of all $f_{\mathbf{l,i}}$, this allows to pose an efficient and consistent on-the-fly scheme.

Conceptually, we start with a SG $f_L$ with undefined values for the $f_{\mathbf{l,i}}$. When a query arrives, we replace those undefined values with the correct ones which are needed to interpolate at the query points. Only for those grid points, we actually run simulations of the HFM. When the next query arrives, we do the same again, of course, replacing only undefined values. In practice, we employ a data structure which only carries the defined grid points, expanding when new grid points and simulation data needs to be added. For the chosen set of basis functions, this can be done in a self-consistent way such that the on-the-fly version of $f_L(\mathbf{x})$ always gives the exact same value at the query points as the corresponding full SG and is never more expensive in terms of high-fidelity simulations. Figure \ref{fig:1D_basis}c illustrates the effect of the on-the-fly refined grid in two dimensions, with an arbitrary solution trajectory depicted in blue. Only the for the shown grid points the high-fidelity function $f$ needs to be evaluated.

The idea is further enhanced by adapting a multilevel approach originally proposed for the solution of nonlinear partial differential equations\cite{bank1982analysis}.
In our adaption, we  start with an initial interpolation level $L_0$ and obtain the corresponding solution. We then subsequently increase the level, taking the previously obtained solution as a starting point. This is repeated until convergence is reached.
This is motivated by that an accurate surrogate is not necessary far away from the true solution. With increasing level we converge to the true solution and therefore save unnecessary evaluations far away from the solution. Further, this procedure has the convergence test already built-in. Also, the multilevel scheme pairs well with the noise balancing approach, which we will introduce in the next section. For further details, we refer to our previous publication \cite{hulser2025multilevel}.\\

\subsection{Monte Carlo type models}\label{sec:mc-models}
The above outlined  ML-OTF-SG approach implicitly assumes that we can evaluate the function $f$ without any errors due to the underlying high-fidelity simulation. In our previous work \cite{hulser2025multilevel}, this could safely be assumed because the employed mean field models could be solved with negligible numerical errors. However, for more complex models, simply running highly accurate high-fidelity simulations is easily out of reach. This hold particularly true for Monte-Carlo-type models, where the expectation in Equation \eqref{eq:constitutive-relation-expectation} must be estimated by sampling converging very slowly. Besides the discretization error, which is also present with exact HFM evaluations, the surrogate's accuracy will thus be affected by these additional errors. These errors must be balanced to ensure convergence of the surrogate against the true constitutive function (CF) and an efficient coupling.

As discussed, we assume that the CF value for a given input $X$ is derived from some kind of parametric stochastic model $S(\mathbf{x})$, i.e., $S(\mathbf{x})$ is drawn from a conditional probability distribution $P(S|\mathbf{x})$. We further assume that we have a kind of Monte Carlo simulation which delivers independent samples. When the transformation $T$ in Equation \eqref{eq:constitutive-relation-expectation} is affine, the estimates $\tilde{f}_{\mathbf{l,i}}$ for the true CF values ${f}_{\mathbf{l,i}}$ at the grid points then obey
\begin{equation}
 \tilde{f}_{\mathbf{l,i}}= \frac{1}{N_{\mathbf{l,i}}} \sum\limits_{n=1}^{N_{\mathbf{l,i}}} T(s^{n}_{\mathbf{l,i}} )={f}_{\mathbf{l,i}} + \epsilon_{\mathbf{l,i}}
\end{equation}
where we used the empirical mean to estimate the conditional expectation in \eqref{eq:constitutive-relation-expectation} from $ N_{\mathbf{l,i}}$ samples $\{s^{n}_{\mathbf{l,i}}\}_{n=1}^{\mathbf{l,i}}$. The deviation $\epsilon_{\mathbf{l,i}}$ from the true CF values ${f}_{\mathbf{l,i}}$ is a random noise with zero mean and a statistics which depends on $ N_{\mathbf{l,i}}$ and $\mathbf{x}_{\mathbf{l,i}}$.

Let $f_L,\tilde{f}_L$ be the full SG approximation up to level $L$ using $f_{\mathbf{l,i}}$ and $ \tilde{f}_{\mathbf{l,i}}$. In order to control the error introduced by the noise, we consider the expected squared 2-norm error $E(||f-\tilde{f_L}||^2_2)$ averaging with respect to the stochastic nature of the $\tilde{f}_{\mathbf{l,i}}$. For the assumed affine transformation and independent samples, this error measure can be decomposed as
\begin{align}\label{eq:total-error}
E(||f-\tilde{f_L}||^2_2)=||f-f_L||_2^2 + E(||f_L-\tilde f_L||_2^2) \nonumber\\
=|| f-f_L||^2_2  + \sum_{\mathbf{l},|\mathbf{l}|_1\leq L} \sum_{\mathbf{i} \in I_\mathbf{l}} z_{{\mathbf{l,i}},L} \text{Var}( \tilde f_{\mathbf{l,i}})
\end{align}
where the coefficients $z_{{\mathbf{l,i}},L}>0$ can be precomputed without any information on the function f or the stochastic model for its pointwise estimation. For the derivation, we refer to the App. \ref{app:error-splitting}. This way, we have split the error into two independent parts, the noise free discretization error $||f-f_L||_2^2$ and a noise induced error $E(||f_L-\tilde f_L||_2^2)= \sum_{\mathbf{l},|\mathbf{l}|_1\leq L} \sum_{\mathbf{i} \in I_\mathbf{l}} z_{{\mathbf{l,i}},L} \text{Var}( \tilde f_{\mathbf{l,i}})$ which is zero in the noise free case.
The interpolation error  $||f-f_L||_2^2$ is controlled by the interpolation level $L$ and completely independent of the stochastic nature of the function value approximations. In contrast,  noise induced error $E(||f_L-\tilde f_L||_2^2)$ can be calculated from the variances of the estimates $\tilde  f_{\mathbf{l,i}}$. The estimate $\tilde  f_{\mathbf{l,i}}$ is just an empirical mean of $T(S(\mathbf{x}_{\mathbf{l,i}}))$ and its variance is simply $\text{Var}(T(S(\mathbf{x}_{\mathbf{l,i}})))/N_{\mathbf{l,i}}$.

To exploit this decomposition to control the sampling effort $N_{\mathbf{l,i}}$ per grid point, we assume that there exist an upper bound $V$ for the variance of the transformed stochastic model $T(S(\mathbf{x}))$, which is independent of $\mathbf{x}$. The noise induced error then possesses an upper bound
\begin{equation}\label{eq:sampling-error-term}
 E(||f_L-\tilde f_L||_2^2) \leq  V \sum_{\mathbf{l},|\mathbf{l}|_1\leq L} \sum_{\mathbf{i} \in I_\mathbf{l}} z_{{\mathbf{l,i}},L} N_{\mathbf{l,i}}^{-1}.
\end{equation}
We now require that this bound is smaller than some predefined value $\epsilon_L^2>0$ and minimize the total cost under this constraint, i.e., we minimize the total number of samples $\sum_{\mathbf{l},|\mathbf{l}|_1\leq L} \sum_{\mathbf{i} \in I_\mathbf{l}}N_{\mathbf{l,i}}$ allowing independent variation of the different $N_{\mathbf{l,i}}$.
For a SG of level $L$, we find for the sampling effort per grid point
\begin{equation}\label{eq:optimal-sampling}
N_{\mathbf{l,i},L} \geq \frac{V}{\epsilon_L^2} Z_{\mathbf{l,i},L}
\end{equation}
where we introduced an index $L$ to indicate the dependence on the maximum level $L$ and used the abbreviation $Z_{\mathbf{l,i},L}:=\sqrt{z_{\mathbf{l,i},L}} \sum_{\mathbf{l},|\mathbf{l}|_1\leq L} \sum_{\mathbf{i} \in I_\mathbf{l}} \sqrt{z_{{\mathbf{l,i}},L}}$.
Choosing the numbers of samples according to Equation \eqref{eq:optimal-sampling} will lead to a converging scheme for any choice of $\epsilon_L$ which decays to zero for increasing $L$.

For further insight on how to choose $\epsilon_L$, we investigate the decomposition Equation \eqref{eq:total-error}. This implies that a fine SG with low discretization error $||f-f_L||_2$ combined with only inaccurate sampling, i.e., large $\epsilon_L$, will be a waste of grid points because the sampling error will dominate the total approximation error. In the opposite extreme, extremely well sampled data will have little impact on an only coarse SG because the total error is dominated by the discretization error. We thus expect the best balance when sampling and discretization error are in the same order. Now, $||f-f_L||_2$ is unknown, especially for the ML-OTF-SG strategy, where we do not construct a full SG. We therefore need a reasonable approximation, and an obvious choice would be the bound Equation \eqref{eq:SG_error_bound}. Typically, this tends to largely overestimate the error at low levels and only becomes representative of the true error in the limit of large levels. For low $L$, we would therefore have a too high noise error, making the error balancing only useful for high levels, a limit we will likely never reach in higher dimensions. However, the bound \eqref{eq:SG_error_bound} implies that error decay beyond second order, i.e., $||f-f_L||_2=O(2^{-2L})$, will be a coincidence and second order decay can only be expected in the asymptotic limit of large $L$. We therefore
choose a first order decay for $\epsilon_L$, i.e.
\begin{equation}
 \epsilon_L= C_{\epsilon} 2^{-L}
\end{equation}
which is a good compromise between decay of the sampling error and increase of sampling effort with $L$.

The above analysis leaves us with two constants controlling the sampling effort and the error balancing, $C_{\epsilon}$ and the bound for the variance $V$. The later can be eliminated by turning Equation \eqref{eq:optimal-sampling} into a bound for the variance of the estimate
\begin{equation}\label{eq:Var_j-total-final}
 \mathrm{Var}(\Tilde{f}_{\mathbf{l,i},L})  \leq  C_{\epsilon}^2 \frac{1}{Z_{\mathbf{l,i},L}} 2^{-2L},
 \end{equation}
where $ \mathrm{Var}(\Tilde{f}_{\mathbf{l,i},L})$ can be estimated from a few samples on the fly.
The remaining constant $C_{\epsilon}$ is a parameter which can, in principle, be freely chosen. To obtain a value which leads to neither excessive nor overly poor sampling, we follow a heuristic strategy geared toward ML-OTF-SGs. For ML-OTF-SG, we do not need to control the errors on the whole hypercube but only on the subset, which has actually been visited. Therefore, we create a full $L=1$-SG $\bar{f}_1$ and gradually increase $C_{\epsilon}$ until the sampling error is small compared to the norm of the interpolation. We then run a coupled simulation with this SG and obtain rough approximate values $\mathbf{x}_g$ at the relevant spatial points of the continuum simulation, i.e., those spatial points where the CF must be evaluated. For these  $\mathbf{x}_g$, we create reference estimates $f_{\text{R},g}$ of the true CF values from sampling the stochastic model and the corresponding values $\bar{f}_1(\mathbf{x}_g)$ provided by the $L=1$-SG. We then obtain $C_{\epsilon}$ by requiring
\begin{equation}\label{eq:choice-sampling-constant}
 \epsilon_{L=1}= \sqrt{\frac{1}{N_{\text{spatial}}} \sum\limits_{g=1}^{N_{\text{spatial}}} \left( f_{\text{R},g} - \bar{f}_1(\mathbf{x}_g) \right)^2 }
\end{equation}
where $N_{\text{spatial}}$ is the number of spatial points employed. The term on the right-hand-side of Equation \eqref{eq:choice-sampling-constant}  is an estimate for a (weighted) 2-norm error on the subdomain of $[-0.5,0.5]^d$ occupied by the solution. This is the equivalent to the above considered $||f-f_{L=1}||_2$ defined on the whole domain $[-0.5,0.5]^d$ and  Equation \eqref{eq:choice-sampling-constant} is a reasonable choice in the ML-OTF-SG context. Note, this procedure can be conducted with only low accuracy, as $\epsilon_L$ will only qualitatively resemble the discretization error. If the number of spatial points $N_{\text{spatial}}$ is large, the empirical mean in Equation \eqref{eq:choice-sampling-constant} can therefore be conducted by with only a small randomly chosen subset of all points and, hence, much less computational costs for sampling reference evaluations.

For simplicity, the above has been derived for affine transformations $T$ (cf. Equation \eqref{eq:constitutive-relation-expectation}). In practice, it might be better to employ more flexible nonlinear transformation, instead. Reasons might be that then the transformed expectation is easier to interpolate or the variance bounds become tighter, leading to less conservative estimates for the number of samples per grid point, Equation \eqref{eq:optimal-sampling}. Such general nonlinear transformations will introduce additional terms into the error decomposition \eqref{eq:total-error} as the estimates $ \tilde{f}_{\mathbf{l,i}}$ would carry a bias additionally to the random noise. Using the rule \eqref{eq:optimal-sampling}, these bias-induced errors however decays with $L$ fulfilling a similar, conservative bound as \eqref{eq:sampling-error-term}, this is shown in the App. \ref{app:noise-bias}.
It is thus sufficient to control only the noise-induced error, and the above scheme is also valid for general nonlinear transformations  with the adaption that the variances $ \text{Var}\tilde{f}_{\mathbf{l,i}}$ will be estimated using bootstrapping \cite{davison1997bootstrap}.

\section{Test cases: Fixed bed reactor model coupled to kinetic Monte Carlo models of heterogeneous catalysis}\label{sec:results}
\begin{figure}
    \centering
    \includegraphics[width=0.5\textwidth]{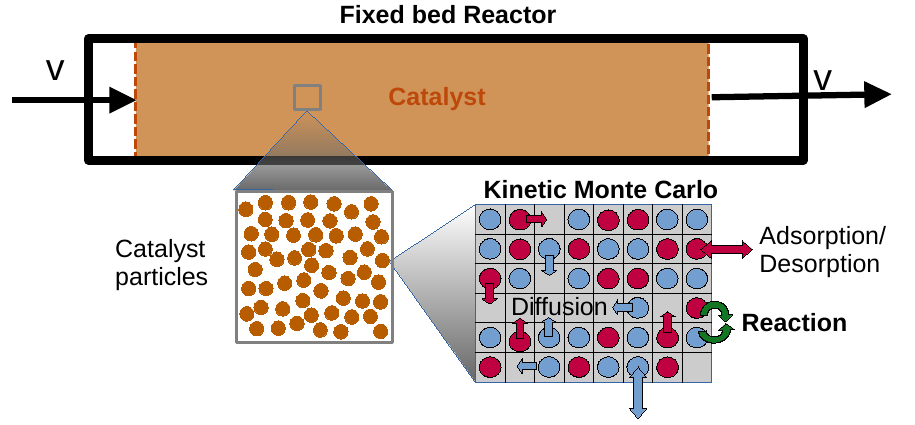}
    \caption{Illustration of multiscale modeling of heterogeneous catalysis in a fixed-bed reactor.}
    \label{fig:multiscale-reactor}
\end{figure}

We demonstrate the approach on a simple, yet realistic model from the field of heterogeneous catalysis, which is the conversion of gaseous chemical species by surface reactions. Within this context, we consider a stationary and isothermal  fixed-bed-reactor, i.e., a tube filled with catalyst particles \cite{fogler1999elements}.
On the macroscopic continuum scale, we employ a coarse-grained description for the transport and conversion of the gaseous species, where one representative volume element contains a large number of catalytic particles. Microscopically, we model the interplay of elementary reactions on the active sites on the particles' surfaces by  high-dimensional Markov jump processes,  which will be simulated using the kinetic Monte Carlo methodology \cite{andersen2019practical,pineda2022kinetic}. 
\textbf{Figure} \ref{fig:multiscale-reactor} illustrates this multiscale model ranging between micro and macro scale.

We assume a straight tube, which can be modeled by a set of one-dimensional convection-diffusion-reaction equations \cite{stynes2018convection}. In particular, we consider
\begin{align}\label{eq:fixed-bed-reactor-model_concentration}
  v \frac{d}{dq} c_\alpha -  D \frac{d^2}{dq^2} c_\alpha =K I(q) \sum\limits_{i}\nu_{i,\alpha} R_i(\mathbf{c}),
\end{align}
which are subject to Dirichlet boundary conditions ($c_\alpha$=const.) at the inlet $(q=0)$ and Neumann boundary conditions $(\frac{d}{d_q} c_\alpha=0)$ at the outlet (at $q=\Lambda=10$\,mm).
In Equation \eqref{eq:fixed-bed-reactor-model_concentration}, $c_\alpha$ denotes the species concentration and $\mathbf{c}$ is the vector of all species concentrations. In all our tests, we assume $v=2.1\times 10^{-3}$\,m/s for the flow velocity of the mixture and $D=2.1\times 10^{-6}$ for the diffusion coefficient which we assume to be the same for all species, for simplicity. The reactions are modeled as a volumetric source term $K I(q)\sum_{i}\nu_{i,\alpha} R_i$, where $\nu_{i,\alpha}$ is the stoichiometric coefficient of the species $\alpha$ in the $i$-th reaction and
$K$ is the amount of active sites per volume. The intrinsic reactivity $R_i$ of the $i$-th reaction is the expected number of events of this reaction per active site and time. This is commonly termed turnover frequency (TOF) and depends also on temperature. $I(q)$ is an indicator function accounting for tubes which are partially filled with inert particles. We assume $I(q)=1$ for $q< \Lambda_C= 8$\,mm and zero for $q\geq \Lambda_C$, i.e., we assume that the last $2$\,mm are filled with inert particles. We discretize the Equation \eqref{eq:fixed-bed-reactor-model_concentration} using central finite differences with $N_\text{dis}=100$ equidistant grid points. To numerically solve the resulting set of nonlinear equations, we employ Powell's hybrid method, as implemented in the GNU Scientific library \cite{powell1970hybrid,gsl}. The employed default parameters for the model and the discretization can be found in Tab. \ref{tab:simulation-parameters}.

The TOF function $R_i(\mathbf{c})$ is an expectation of a stationary Markov jump process, which is defined by a microkinetic reaction mechanism with rate constants parametrically depending on the concentrations \cite{andersen2019practical}. Since the process is high-dimensional, the corresponding master equation is numerically intractable except for  problems with a special structure of the stochastic generator \cite{gelss2016solving,gelss2017nearest,pineda2017beyond,kumar2023probabilistic}. 
The kinetic Monte Carlo (kMC) methodology circumvents this problem by simulating unbiased sample trajectories of the stochastic process. In practice, a kMC model of heterogeneous catalysis works on a lattice of adsorption sites on the surface, where reaction intermediates can bind. A state in this model is then characterized by the occupation of the individual sites, i.e., which intermediate is bound to which site. A kMC simulation evolves this state in a random fashion by executing one reactive event after the other \cite{jansen2012introduction}.
The TOF $R_i$ for a given set of inputs $\mathbf{c}$ is obtained from the stoichiometry and the stationary expected rates of these elementary processes. Estimating such expectations involves creating samples by first running a large number of kMC steps to relax the system to steady state, starting from a predefined state, and a subsequent time averaging over a similar large number of steps.

A characteristic of kinetics is its high degree of nonlinearity, often with variations of several orders of magnitude within narrow ranges of concentrations and temperature.
It is therefore common practice to consider  the logarithms of the rates as functions of the logarithmic pressures and inverse temperature \cite{matera2014predictive,doppel2023efficient,klumpers2023direct}. Also,
kinetic expressions often get a much simpler functional form in this representation and, thus, leading to a potentially simpler interpolation task. However, there is a drawback with such representations, when one has finite noise induced errors. The variance of the logarithm of the rate estimates typically becomes arbitrarily large when the expectation value goes to 0. For the balancing, this would imply that we would have to spend a lot of sampling effort to reduce variances further that are essentially zero, just because the log representation scales them to large values. Employing an affine transformation for the rates has the advantage that such errors stay small and, as the rates enter the coupled problem in a linear fashion, will not affect the solution for the target concentration fields much. At the same time, the interpolation becomes more demanding, because the interpolation error will be controlled by the largest values of the rates. These can be orders of magnitude larger than the values which are taken in the coupled problem. We then might have large relative errors for the targeted concentration field, because, by the interpolation, the high values ``poison'' into the relevant partial pressure and temperatures regimes. To overcome both limitations, we employ an inverse hyperbolic sine transformation, which has been proposed for surrogate modeling of mean field kinetic models \cite{doppel2023efficient}. In detail,  we interpolate $\text{arcsinh}(R_\alpha/R_{\alpha,\text{MTL}})$ as a function of the logarithmic pressure and inverse temperature (both subject to a componentwise affine transformation which maps relevant the ranges to $[-0.5,0.5]$. This transformation behaves close to linear for $| R_\alpha| < R_{\alpha,\text{ref.}}$ and close to logarithmic for  $| R_\alpha| > R_{\alpha,\text{ref.}}$. Values much larger than $R_{\alpha,\text{ref.}}$ are thus effectively lowered relative to the relevant range with smaller values. The reference rate $R_{\alpha,\text{ref.}}$ defines therefore the target range for the error, i.e., rates in the same order of magnitude should lead to significant deviations of the concentration fields from the inlet condition. A reasonable value for $R_{\alpha,\text{ref.}}$  can be found by exploiting that concentrations can not be negative. We therefore determine $R_{\alpha,\text{ref.}}$, by solving the continuum model \eqref{eq:fixed-bed-reactor-model_concentration} with spatially constant values for all $R_{\alpha,\text{ref.}}$ (of course respecting stoichiometry). We vary them until the concentration of one species drops to zero and take those values as reference rates (as the problem is linear, this requires actually only a single simulation). The argument behind this choice is that rates which lead to significant changes in the concentration must be in the order of magnitude of these values, while at the same time values which are much larger can not appear because they would lead to locally negative concentrations.

In this study, we will investigate three different kinetic models. One is an analytical model for testing purposes, where we add artificial noise to mimic real kMC simulations. The other two are so-called first principles kinetic Monte Carlo models, i.e., the reaction mechanism and the corresponding rate constant expressions have been obtained from first principles quantum-chemical simulations \cite{andersen2019practical}.
For the actual kMC simulations, we employ the {\sc kmos3} package  \url{https://kmos3.org} \cite{hoffmann2014kmos},
which is a code generator producing highly efficient model specific FORTRAN code. 

Actual computations have been conducted using a single computing node with 72 Intel Xeon IceLake Platinum 8360Y cores. In every query of the nonlinear solver to the ML-OTF-SG, the required additional kMC samples are obtained by running single core kMC simulations parallelized over the available cores.

\begin{table}[htbp]
    \centering
    \begin{tabularx}{0.5\textwidth}{|l| X| l|}
    \hline
        Parameter &Description &Value \\
        \hline
        $\Lambda$ & Fixed bed reactor length & 1 cm \\
        $v$ & Inflow velocity & $2.1 \cdot 10^{-3}$ m/s \\
        $D$ & Diffusion coefficient & $2.1 \cdot 10^{-5} \, \text{m}^2/\text{s}$ \\
        $\Lambda_C$  & Length of catalytic area & 0.8 mm\\
         $N_\text{dis}$  & Spatial discretization & 100\\
         \hline
    \end{tabularx}
    \caption{simulation parameters}
    \label{tab:simulation-parameters}
\end{table}

\subsection{Semi-analytical test: Water gas shift on copper based catalysts}\label{sec:result-analytical}

\begin{figure}
		\centering
			\includegraphics[width=0.5\textwidth]{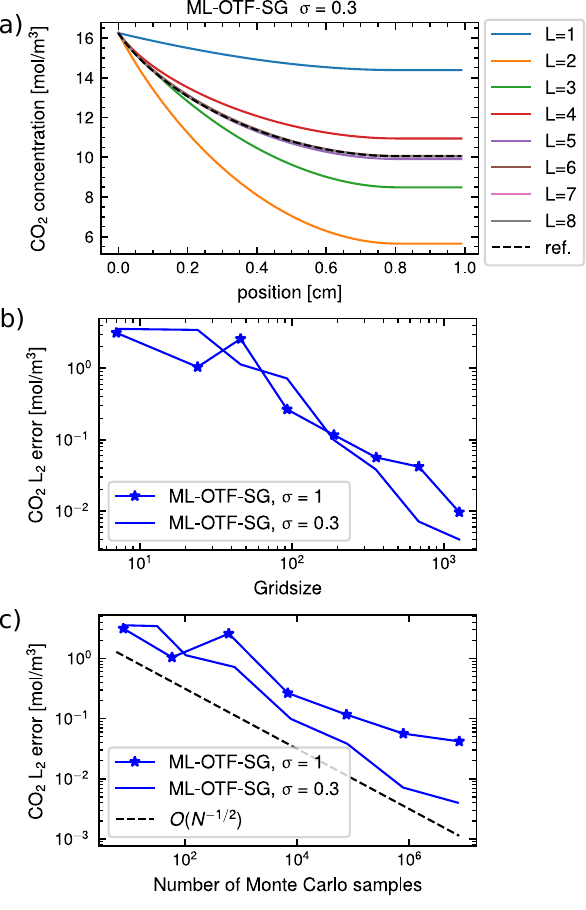}
		\caption{a) Solutions for \ch{CO2} concentration of the coupled system \eqref{eq:fixed-bed-reactor-model_concentration} with the analytical reverse-water-gas-shift model Equation \eqref{eq:rwgs-R} and different level $L$ of on-the-fly interpolation for a noise STD $\sigma = 0.3$. Reference solution from direct coupling without interpolation is shown in the black dotted line. b) $L_2$ error of the noise balanced ML-OTF-SG scheme over grid size of the sparse grid for noise levels $\sigma=0.3$ and $\sigma=1$.
        c) $L_2$ error of the noise balanced ML-OTF-SG scheme over sum of samples $\sum_j N_j$ for noise levels $\sigma=0.3$ and $\sigma=1$ as well as theoretical limiting $O(N^{-1/2})$ scaling for Monte Carlo methods.}
		\label{fig:rwgs_balanced}
\end{figure}

As an introductory case, we consider an analytical (empirical) reverse water gas shift model for a Cu/ZnO catalyst from literature \cite{bussche1996steady} with constants corresponding to \cite{brosigke2022closer}. The reaction equation for the reverse water gas shift is
\begin{equation}
 \mathrm{ CO_2 + H_2 \rightleftharpoons CO + H_2O},
\end{equation}
i.e., stoichiometric coefficients have the value $-1$ for CO$_2$ and H$_2$ and the value $1$ for CO and H$_2$O. The employed model is directly formulated for the expected rate $R_{\text{RWGS}}(\mathbf{c})$ and carries no noise (cf. App. \ref{app::WGS_CuZnO}). To mimic the behavior which we expect for a kMC model, we create a stochastic model by adding relative Gaussian noise to this analytical expression, i.e.,  $R_{\text{RWGS,stoch.}}(\mathbf{c})=(1+\eta)R_{\text{RWGS}}(\mathbf{c})$ where $\eta$ is a Gaussian noise with zero mean and standard deviation (STD) of $\sigma$. This mimics the noise that is to be expected in a kMC model, where low and high TOFs will have a low and high (absolute) STD, respectively.

This artificial stochastic model serves for benchmarking, because we have a closed expression for the conditional expectation. Thereby, we can afford reference solutions which are only affected by the numerical errors due to the discretization of Equation \eqref{eq:fixed-bed-reactor-model_concentration}. As a test case, we consider a temperature $T=740$\,K and a mixture at the inlet with partial pressures of $1$\,bar and $6$\,bar for CO$_2$ and H$_2$, respectively, and $10^{-5}$\,bar for CO and H$_2$O. As an empirical model, reactivities are normalized to the mass of catalyst, and we employ a catalyst loading of $K=1\si{kg_{cat.}m^{-3}}$. 
The partial pressure ranges for interpolation are $[10^{-5},1]$\,bar for H$_2$, CO, and H$_2$O and $[10^{-5},6]$ for CO$_2$. To illustrate the effect of differently strong noise, we consider the two differently strong noises: i) medium strong noise with $\sigma=0.3$ and ii) strong noise with $\sigma=1.0$

For the case $\sigma=0.3$, the solutions for the \ch{CO2} concentration along the reactor are shown in \textbf{Figure} \ref{fig:rwgs_balanced}a for different maximal level $L$ in the ML-OTF-SG (differently colored solid lines) and for the reference solution (dashed black line).
The other species as well as the high noise case ($\sigma=1$) show a comparable behavior. For level $L=1,2$, we still find significant, but only quantitative deviations from the reference. For higher levels, the curves converge with increasing $L$ against the reference solution, with the solution for $L=3$ already being in the range $10$\,\% error and much lower errors for higher levels.

To illustrate the convergence with the grid refinement, Figure \ref{fig:rwgs_balanced}b shows the normalized L$_2$-error of the \ch{CO2} concentration of the ML-OTF-SG as function of
the grid size of the SG for both noise cases. As expected from Figure \ref{fig:rwgs_balanced}a, we see a fast convergence with the grid size for $\sigma=0.3$, but also for the high noise case with $\sigma=1$, very low errors with modest grid sizes. Both cases behave very similar, showing an close to first order convergence with the number of grid points. This similar behavior is expected by the sampling control strategy, but fluctuations in the order of the error are to be expected due to the fluctuating stochastic error.

To illustrate the convergence with the invested Monte Carlo sampling effort,
Figure \ref{fig:rwgs_balanced}c shows the same errors in dependence on the total number of samples $N=\sum_{\mathbf{l,i}} N_{\mathbf{l,i},L}$. For comparison, we also show a $O(N^{-1/2})$ curve, which is the best asymptotic scaling which can be achieved on basis of Monte Carlo data \cite{heinrich1999monte}. As expected, both cases show convergence, with the lower-noise case showing lower errors at larger numbers of samples. The lower-noise case also follows the $O(N^{-1/2})$ asymptotics, whereas the high-noise curve shows slower convergence. The latter is to be expected as a $O(N^{-1/2})$ convergence can only be achieved for properly coupled Monte Carlo samples \cite{heinrich1999monte} in contrast to the employed independent samples mimicking a black-box Monte Carlo model. The close to ideal convergence of the case with $\sigma=0.3$ is therefore a result of the smaller noise resulting in a convergence behavior which is closer to the deterministic case.

\subsection{CO-oxidation on \ch{RuO2}}\label{sec:CO-oxidation-RuO2}

\begin{figure}
		\centering
			\includegraphics[width=0.5\textwidth]{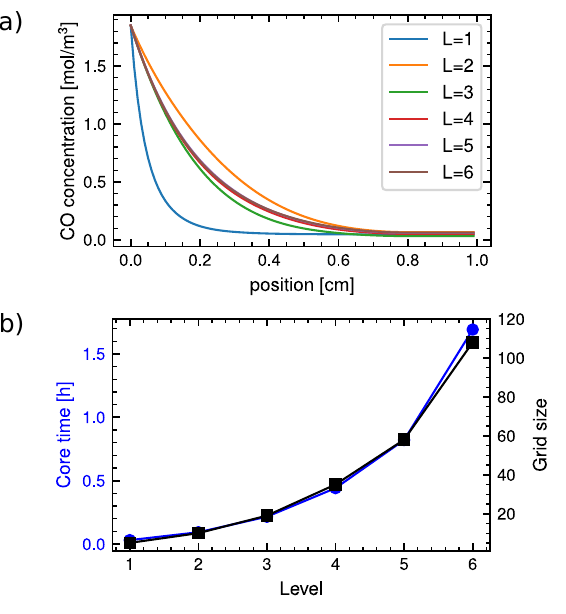}
		\caption{a) Solutions for the \ch{CO} partial pressure of the coupled system \eqref{eq:fixed-bed-reactor-model_concentration} for the \ch{RuO2} \ch{CO}-oxidation model (Section \ref{sec:CO-oxidation-RuO2}) and different levels $L$ of on-the-fly interpolation. b) Approximate computational cost (blues circles left axis) and number of grid points (black squares right axis)
		}
		\label{fig:RuO2}
\end{figure}
As the first realistic example,  we consider the \ch{CO} oxidation reaction
\begin{equation}
 \mathrm{ CO + \frac{1}{2}O_2 \rightleftharpoons CO_2},
\end{equation}
particularly the 1p-kMC model for the \ch{RuO2}(110) surface introduced by Reuter and Scheffler \cite{reuter2006first}. 
This has been a showcase of many theoretical and computational studies, e.g. \cite{pogodin2014more,herschlag2015consistent,gelss2016solving,hess2017probing,dortaj2023efficient},
 particularly in the context of coupling kMC and reactive transport simulations \cite{matera2010transport,mei2011effects,sutton2018electrons,bracconi2020training}.
The reaction is effectively irreversible for realistic partial pressures and the \ch{RuO2}(110) surface, because the CO$_2$ adsorption has a very high barrier. This has the effect that the TOF for this reaction only depends on the partial pressures of \ch{CO} and \ch{O2}, but not on the partial pressure of \ch{CO2}. From the surrogate point of view, the model is thus just two-dimensional.

For our test, we consider a mixture of \ch{CO}, \ch{O2} and \ch{CO2} with partial pressures of $0.1$\,bar for \ch{CO} and \ch{CO2} and $0.5$\,bar for \ch{O2} at the inlet and a temperature of $T=650K$. For the catalyst loading, we employ here $K=1 \frac{mol}{m^3}$. The interpolation is conducted on the range $[1\times 10^{-5},1]$\,bar for both partial pressures. For the kMC simulation, we consider a two-dimensional periodic lattice of $20\times 20$ surface unit cells. We employ $10^6$ kMC steps to relax the system to steady state, and TOF samples are created by time-averaging over the subsequent $10^6$ kMC steps.

\begin{figure} 
		\centering
			\includegraphics[width=0.5\textwidth]{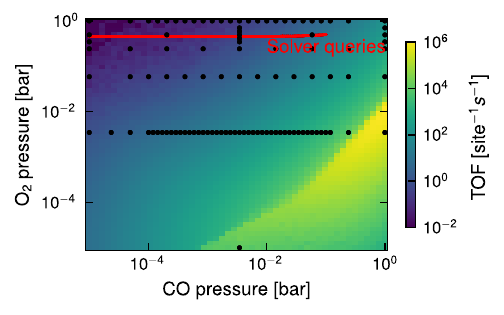}
		\caption{Turnover frequency of the \ch{RuO2} model over $p_{\ch{CO}}$, $p_{\ch{O2}}$ with $p_{\ch{O2}},p_{\ch{CO}} \in [1\times 10^{-5},1] $bar and $T=650\si{K}$ averaged over 10 samples per evaluation point. ML-OTF-SG solver query points are depicted in red, the corresponding grid points as black dots.}
		\label{fig:TOF-RuO2}
\end{figure}

\textbf{Figure} \ref{fig:RuO2}a shows the solutions of the ML-OTF-SG for the \ch{CO}-concentration over the reactor coordinate $q$ for different maximum levels $L$. The solutions converge quickly with the level, where levels $L \geq 3$ can hardly be distinguished anymore. As for the analytical test case, the other species concentrations show the same behavior. Irrespective of the employed level, the computational costs stay modest as can be seen in Figure \ref{fig:RuO2}b which shows the total number of core hours spend on kMC sampling in dependence of the level. Even for the highest level $L=6$, less than two core hours are required, making the simulation very affordable. The very low costs for this problem can, to some extent, explained with the very good sampling properties of this 1p-kMC model. The variances of the TOF samples are small, especially when compared to other 1p-kMC models. The very low variance also explains the modest increase of the costs with the levels. This is reflected by the grid size in dependence of the level $L$ also shown in Figure \ref{fig:RuO2}b. This is almost perfectly proportional to the core time, i.e., the number of samples can stay at their initial value when increasing the level.

Another characteristic of this model is its very high peak TOF in the considered range of partial pressures, as well as its high degree of nonlinearity. \textbf{Figure} \ref{fig:TOF-RuO2} display the TOF as a function of the \ch{CO} and \ch{O2} partial pressures, $p_\mathrm{CO}$ and $p_\mathrm{O_2}$, respectively. We see a rim of very high TOF going from low values for $p_\mathrm{CO}$ and $p_\mathrm{O_2}$ to high values. Close to this rim, the TOF varies rapidly, making this region difficult to interpolate. However, having an excess of the \ch{O2}-species, the simulation is done at regions with only moderate TOF variations. Also, the highest TOFs are several orders of magnitude higher than those taken in the coupled simulation. This shows the benefits of the nonlinear transform. Without it these high values would lead to tremendous errors of the interpolation for the relevant partial pressures, at least at lower levels. Also, the TOF is rather simple to interpolate close to the relevant partial pressures and the local interpolation errors are small for higher levels. Our local heuristics for estimating a good error bound (cf. Sec. \ref{sec:mc-models}) would be very beneficial for models with stronger noise, as a global alternative would be affected by the large interpolation error for partial pressures, which are irrelevant for the considered problem setting.

\subsection{Water gas shift on Pt(111)}\label{sec:WGS}

\begin{table}[htbp]
    \centering
    \begin{tabular}{|l|l|l|}
        \hline
        Parameter &Inlet value &Input range\\
        \hline
        $\mathrm{p}_{\ch{CO2}}$ & $1\times 10^{-5}$\,bar  &   $ [1\times 10^{-5}, 0.2]$\,bar  \\
        $\mathrm{p}_{\ch{CO}}$ & $0.2$\,bar & $ [1\times 10^{-4}, 0.3 ] $\,bar \\
        $\mathrm{p}_{\ch{H2O}}$ & $0.8$\,bar & $ [0.5, 1]$\,bar \\
        $\mathrm{p}_{\ch{H2}}$ & $1\times 10^{-5}$\,bar  & $[1\times 10^{-5}, 0.2]$\, bar \\
                T & 710 K & [560,760] K\\
        \hline
    \end{tabular}
    \caption{Species, initial compositions and input ranges water gas shift on \ch{Pt}(111)}
    \label{tab:species-compositions-WGS}
\end{table}

The last example is again the water gas shift reaction $\ch{CO + H2O <=> CO2 + H2}$, but now we consider the Pt(111) surface and employ a simplified version of the 1p-kMC model introduced by Stamatakis and Vlachos \cite{stamatakis2011graph}. Details of the model can be found in App. \ref{app::WGS}. We employ  a catalyst loading of $K=1$\,mol m$^{-3}$ and a mixture of \ch{CO,CO2,H2} and \ch{H2O}, for which we choose $p_\mathrm{CO}=0.2$\,bar,  $p_{\mathrm{H_2O}}=0.8$\,bar, and $p_\mathrm{CO_2}=p_\mathrm{H_2}=1\times 10^{-5}$\,bar at the inlet. For the interpolation, we also include the parametric dependence of the TOF on the (inverse) temperature additionally to dependence on the partial pressures. The corresponding interpolation ranges can be found in Tab. \ref{tab:species-compositions-WGS}. For all grid points, we employed $2 \times 10^6$ relaxation and sampling steps in the kMC simulation to create one sample.

\begin{figure}
		\centering
			\includegraphics[width=0.5\textwidth]{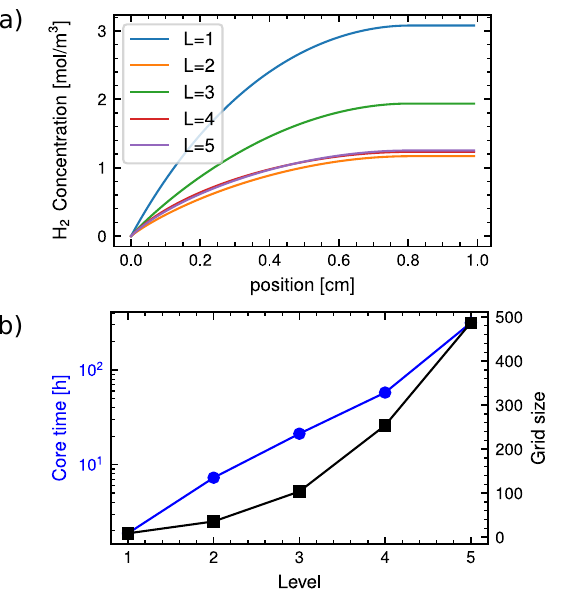}
		\caption{a) Solutions for the \ch{H2} partial pressure of the coupled system \eqref{eq:fixed-bed-reactor-model_concentration} with the WGS model (cf. Sec. \ref{sec:WGS}) and different levels $L$ of on-the-fly interpolation for $T=700K$. b) Computational cost and grid size vs. level $L$.
		}
		\label{fig:WGS}
\end{figure}
As a first test case, we consider a reactor temperature of $700$\,K. \textbf{Figure} \ref{fig:WGS}a shows the obtained partial pressures for \ch{H_2} over spatial position for different levels of SGI. The solution converged with increasing level, where the results for levels $L=4$ and $L=5$ are almost indistinguishable. In Figure \ref{fig:WGS}b we show the computational cost needed for the 1p-kMC sampling in this setup. While lower levels only require a few core hours, the cost increases exponentially with increasing level, with $246$ core hours for level $L=5$. This amount of core time is easily feasible on modern computing architectures. Nevertheless, the cost increases exponentially with increasing level, limiting the possible maximum level.
The computational cost is now significantly higher than for the CO oxidation model in the previous section (cf. Figure \ref{fig:RuO2}b). This is primarily caused by higher noise in the underlying kMC model and the higher dimensionality resulting in a higher number of grid points.

\begin{figure}
		\centering
			\includegraphics[width=0.5\textwidth]{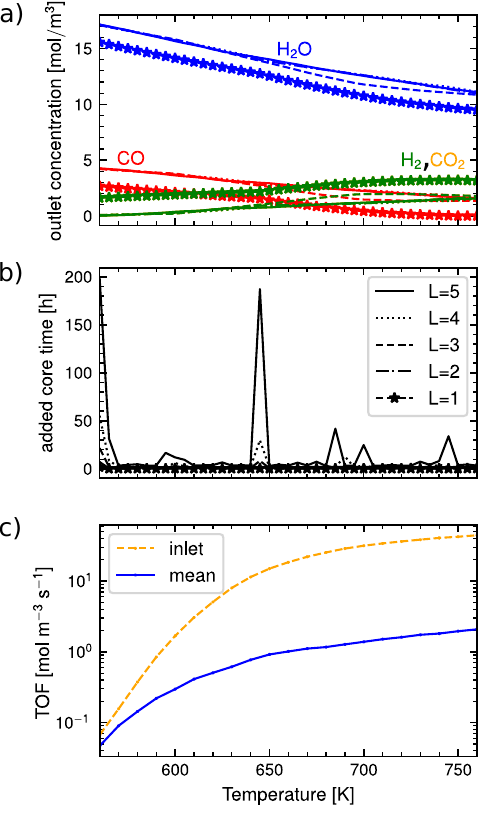}
		\caption{a) Outlet concentrations over temperature $T$ for the coupled system \eqref{eq:fixed-bed-reactor-model_concentration} with the WGS model \ref{sec:WGS} for different levels of interpolation. b) Added computational cost for the noise balanced ML-OTF-SG solver. c) Mean Turnover frequency (TOF) and inlet TOF over temperature for the level 5 case.}
		\label{fig:wgs-T-scan}
\end{figure}
An important advantage of traditional surrogate models is their reusability, i.e., once created, they can be used in many different large scale simulations without extra computational burden by microscale simulations. For our ML-OTF-SG strategy, this is only partly true, as the surrogate is constructed during the simulation. A new simulation can reuse the existing data but might need to create new data in other simulations, of course. To address this point, we scan over different values of the temperature for the range $T\in [560,760]K$, which is a typical experimental procedure in heterogeneous catalysis. Starting at the lowest temperature we increase the temperature in steps of $5$\,K always recycling the already existing 1p-kMC data. \textbf{Figure} \ref{fig:wgs-T-scan}a shows the obtained outlet concentrations for the different species with different levels of interpolation. For levels  $L>1$, all simulation results show the same behavior, with high quantitative agreement becoming indistinguishable for higher levels. Only for $L=1$, we find a constant shift throughout the whole temperature range. The errors are of the same magnitude for all species but of different sign. That is a consequence of the stoichiometry and the conservative nature of the transport terms in the continuum model \eqref{eq:fixed-bed-reactor-model_concentration}. This is also why the \ch{CO2} is invisible as \ch{CO2} and \ch{H2} have the same inlet concentration and the same stoichiometric coefficient.

The computational cost per new temperature is shown in Figure \ref{fig:wgs-T-scan}b.  The ML-OTF-SG only adds evaluations where it needs it for simulation, resulting in spikes of the cost for certain temperature values where new grid points have to be introduced. Level 5 interpolation adds only on two scan steps more than a hundred core hours, and on most steps only a few hours, the lower levels need significantly less. The total cost is a few hundred core hours.

With increasing temperature, we obtain an increase in \ch{CO2} and \ch{H2} concentration while \ch{H2O} and \ch{CO} decrease. This is reflected by an increase in the spatially averaged mean TOF, which is shown in  Figure \ref{fig:wgs-T-scan}c for $L=5$. Also shown is the TOF at the inlet, which in the case of negligible concentration variations should agree with the mean TOF.  We observe that the obtained mean TOF is significantly lower than the inlet TOF by up to a factor of $30$. This reflects on the one hand that we have non-negligible mass transfer effects and on the other hand the high nonlinearity of the problem. Concentrations vary not very much between inlet and outlet, but the TOF shows variations by orders of magnitude.
\section{Discussion}\label{sec:Discussion}

The proposed noise balancing approach was shown to be able to extend the multilevel on-the-fly sparse grid (ML-OTF-SG) approach to a heterogeneous multiscale modeling setting where a continuum model is coupled with a microscopic high-fidelity Monte-Carlo-like model. We showcased the method on a set of problems from heterogeneous catalysis, where the microscopic models are (first principles) kinetic Monte Carlo (1p-kMC) simulations. These are very demanding  examples for surrogate modeling due to their typical high nonlinearity and potential high intrinsic dimensionality, combined with an often bad sampling behavior. The efficiency of the proposed approach manifests in these showcases. The approach delivered highly accurate results at modest computational costs, even for the most demanding case in subsection \ref{sec:WGS}. To our knowledge, a coupling between reactive transport and 1p-kMC models of this complexity has not been achieved so far, especially not with the achieved accuracy.

Besides efficiency, the proposed scheme possesses some properties allowing a convenient usage. Error balancing ensures the convergence against the true expected response of the Monte Carlo model. By construction, the approach also remains the self-consistency of the original ML-OTF-SG approach, i.e., in the coupled simulation, the surrogate appears as an a priori fixed function for the nonlinear solver even though it is constructed on the fly with noisy evaluations. This is the setting nonlinear solvers have been designed for, and the approach can therefore be used with most solvers in a black box fashion, avoiding potential numerical issues resulting from a non-self-consistent on-the-fly surrogate \cite{tynes2025will}. Finally, the approach relies on only a single hyperparameter, the maximum level $L$ of the interpolation. This controls the accuracy as well as the whole surrogate construction process, including which ``training data'' needs to be generated and how accurately. This makes it convenient to use, without the need for extensive hyperparameter optimization and with built-in ``training set''-design.

Efficiency and practicability have been achieved by a decomposition of the surrogate's  total error into the numerical  error due to SG discretization and the stochastic or sampling error due to noisy Monte Carlo data. An important ingredient is then an estimate for the numerical error, where, for guaranteeing convergence, it is sufficient if it decays monotonically with the maximum level $L$. In practice, the choice of this error model matters, as too small estimates would result in overly accurate (and expensive) sampling per grid point and too large estimates would result in unnecessary many SG points. Using some heuristic arguments, we have proposed a simple scheme to obtain estimates which are somehow in between these extremes and work well for the considered showcases. But, for other cases, this might not be sufficient and improving the proposed scheme will be the subject of future research. Another important ingredient are bounds for the sampling variance at the grid points, which need to be known a priori to ensure self-consistency. We employed the simplest of all possibilities: a common bound for all the domain, which has the advantage that its actual value is not needed in the scheme. If better information on the variances can be obtained, this will lead to a better balanced and more efficient scheme.

Using surrogates of Monte Carlo models is often the key to achieve multiscale models which can be applied to realistic scenarios. Many different surrogate modeling approaches have been proposed for this purpose \cite{ankenman2010stochastic,boukouvala2013surrogate,thakur2022deep,zhu2023stochastic},
but only little has been done on balancing the surrogate's discretization error with the Monte-Carlo sampling error \cite{heinrich2001multilevel,alexanderian2014preconditioned,dopking2018multilevel}.
With a priori known grid points, our error balancing can easily be adapted to almost arbitrary interpolation or regression surrogates, which create linear superpositions of basis functions such as polynomial chaos \cite{alexanderian2014preconditioned}
or kernel regression \cite{hofmann2008kernel}. 
Such an extension is not straightforward for existing on-the-fly surrogates \cite{fritzen2019fly,vandermause2020fly,rocha2021fly,haasdonk2023new,scherding2025adaptive}, 
as these usually do not possess the self-consistency property, and for nonlinear regression models, such as Neural networks.

Especially with respect to the balancing of errors, there is space for improving upon the presented approach. One option would be to complement our on-the-fly strategy with an error adaptive refinement. For error adaptive SG, there already exists a noise balancing approach \cite{dopking2018multilevel} and an incorporation into ML-OTF-SG would be rather straightforward. However, this approach is based on a very different reasoning to determine the amount of samples per grid point, which might therefore differ significantly. It is open whether adaptivity would improve upon our non-adaptive approach. Another interesting direction would be a tighter integration into the multiscale modeling process. We have considered the global error of the surrogate on its own high-dimensional domain for our analysis. This makes the approach very flexible and applicable in very different scenarios. However, the central objective is the outcome of the coupled simulation - in our examples the concentration fields, and we have observed that even rather coarse SG surrogates already provide a reasonable accuracy. Extending our error considerations to the coupled problem is thus desirable also to have a criterion at which level $L$ we can truncate. This will necessarily be more problem-specific and requires quantifying the impact of the surrogate's error on the multiscale simulation results.

Concerning the application in the field of heterogeneous catalysis, future steps from our side will include the integration of the ML-OTF-SG with recently developed reduced order Computational Fluid Dynamics approaches \cite{matera2023reduced,qureshi2025reduced} 
to address modern operando experimentation \cite{chen2024advances,dou2017operando,grajciar2018towards}.
These experiments are typically highly affected by macroscopic transport while delivering atomic scale information of the catalyst under operation - an optimal playground for multiscale modeling. 
In this context, it would also be interesting to compare ML-OTF-SG with other on-the-fly strategies \cite{fritzen2019fly,vandermause2020fly,rocha2021fly,haasdonk2023new,scherding2025adaptive}.
These are not self-consistent and also have no error balancing but have shown to work in different scenarios.

\section{Conclusions}
We have presented a novel approach to heterogeneous multiscale modeling, where a Monte Carlo model provides the material response for a continuum model. For this, we introduced a noise-error balancing strategy into our recently developed multilevel on-the-fly sparse grid approach \cite{hulser2025multilevel}, which has originally been developed for purely deterministic models. The approach possesses a unique combination of features making it particularly efficient and convenient to use. The combination of sparse grids with an on-the-fly construction allows the application in high-dimensional settings when the material response depends on a large number macroscopic variables. Noise balancing guarantees convergence as well as quasi-optimal distribution of computational resources to different Monte Carlo tasks. As the original approach, a single hyperparameter, with the meaning of accuracy,  controls the whole process including the ``training set design'' and the amount of Monte Carlo sampling for each data point. It is further designed to work robustly with black-box nonlinear solvers and black-box high-fidelity simulations with controllable noise. We have demonstrated the approach on problems from heterogeneous catalysis, where the approach demonstrated its efficiency achieving highly accurate results with modest computational footprints.

\medskip
\textbf{Acknowledgements} \par 
Funded by the Deutsche Forschungsgemeinschaft (DFG, German Research Foundation) - project no. 467076359 and under Germany’s Excellence Strategy-EXC 2008-390540038-UniSysCat. We want to thank Martin Deimel and Michail Stamatakis for their help with the implementation of the kinetic Monte Carlo models.

\appendix
\section{Error Balancing}

We consider the interpolation domain $\Omega=[-0.5,0.5]^d$ and functions $f:\Omega \mapsto \mathbb{R}$ living in a Hilbert space with the scalar product
\begin{equation}
 \langle a,b \rangle:=\int\limits_{[-0.5,0.5]^d} a(\mathbf{x})b(\mathbf{x}) dx^d
\end{equation}
and the corresponding norm $|| a||_2=\sqrt{ \langle a,a \rangle}$.

We consider the interpolation of the function $f$ using the sparse grid (SG) basis and grid points introduced in section Sec. \ref{sec:multilevel-on-the-fly-sparse-grids}.
For any finite level $L$, the number of basis functions $M_L$ is finite and only depends on $L$ and the dimensionality $d$. For brevity, we therefore introduce integer indices $i,j,k\ldots$ for numbering the basis functions. That is, we denote a basis function by $\phi_k := \phi_{\mathbf{l}_k, \mathbf{j}_k}$ where $\phi_{\mathbf{l}_k, \mathbf{j}_k}$ are the SG basis functions defined in Sec. \ref{sec:multilevel-on-the-fly-sparse-grids}. For simplicity, we employ the convention that if $j > i$, then $|\mathbf{l}_j|_1\geq |\mathbf{l}_i|_1$. A SG interpolant  $f_L=\sum_{i=1}^{M_L} \nu_{i,L} \phi_i$ with maximum level $L$ is uniquely determined by the interpolation condition $f_j:=f(\mathbf{x}_j)=f_L(\mathbf{x}_j)$, leading to a linear system for the surpluses $\nu_{i,L}$. Formally solving this linear system, we can introduce the Lagrange basis functions
\begin{equation}\label{eq:LagrangeBasis}
 \psi_{i,L}=\sum\limits_{j=1}^{M_L} A_{ij,L}\phi_i,
\end{equation}
where the inverse $A^{-1}_L$ of $A$ has the entries $(A^{-1}_L)_{ij}= \phi_i(\mathbf{x}_j)$. Using the Lagrange basis the SG interpolant is given by
\begin{equation}\label{eq:SGInterpolantLagrange}
 f_L= \sum\limits_{i=1}^{M_L} f_i \psi_{i,L}.
\end{equation}
Note that, the SG basis functions $\phi_i$ are independent of the maximum level $L$, but the matrix $A_L$ and the Lagrange basis functions depend on $L$.
As our SG interpolants are exact for a constant function for every $L$, the Lagrange basis is a partition of unity
 \begin{equation}\label{eq:PartitionOfUnity}
 \sum\limits_{i=1}^{M_L} \psi_{i,L}(\mathbf{x}) =1 ,~  \forall \mathbf{x} \in \Omega.
 \end{equation}
We further have
\begin{equation}\label{eq:LowerBoundNormLagrangeBasis}
 \sum\limits_{i=1}^{M_L} \|\psi_{i,L}\|_2 \geq \|  \sum\limits_{i=1}^{M_L} \psi_{i,L} ||_2 = \|1\|_2=1
\end{equation}
where we employed triangle inequality and that $\int_\Omega 1 dx^d=1$.
\subsection{Noise and bias for interpolations with Monte Carlo estimates}
We consider the case, when the function $f$ is given by a transform of a conditional expectation of an underlying (microscopic) stochastic model $S$ (cf. Sec.  \ref{sec:methods}), i.e.
\begin{equation}
 f(\mathbf{x})= T( E(S|\mathbf{x}) ).
\end{equation}
We consider the case when the expectation $E(S|\mathbf{x})$ can not be calculated by analytical or deterministic numerical methods. But, we assume that we have some kind of Monte Carlo method which can produce independent samples from the respective conditional distribution. The expectation is then approximated using the empirical mean and employed in the interpolation, where we allow the number of samples per grid point, $N_{i,L}$, to depend on the maximum level $L$ of the SG interpolation.  The function values ${f}_i$ at the grid point $\mathbf{x}_i$ is related to the corresponding approximation $\tilde{f}_{i,L}$ by
\begin{align}
 \tilde{f}_{i,L} &= T\left( \tilde S_{i,L} \right)= f_i + b_{i,L} + \eta_{i,L} \label{eq:ApproximateFunctionValues}\\
& \text{with} \nonumber \\
 {\tilde S}_{i,L} &:= \frac{1}{N_{i,L}} \sum\limits_{n=1}^{N_{i,L}} s^n_i \label{eq:EmpiricalMean}
\end{align}
where $s^n_i$ is a sample of $S$ conditioned on $\mathbf{x}_i$ and $N_{i,L}$ is the number of samples. The deviation consists
of a zero mean random noise $\eta_{i,L}$ and a bias $b_{i,L}=E(T\left( {\tilde S}_{i,L} \right)) - T( E( \tilde S_{i,L} ) )$ where the expectation is taken with respect to the randomness of the samples. Note, that $E( \tilde S_{i,L} )=E(S|\mathbf{x}_i)$.
The delta-theorem provides bounds for the variance of the noise (and therefore the variance of $\tilde{f}_i$) as well as for the bias \cite{dasgupta2008asymptotic}
\begin{align}
 \text{Var}(\eta_{i,L}) &\leq \frac{V_{i}}{N_{i,L}}\leq \frac{V}{N_{i,L}} \label{eq:VarianceBound}\\
 b_i^2\leq \frac{B_i}{N^2_{i,L}}& \leq \frac{B}{N^2_{i,L}}\label{eq:BiasBound}
\end{align}
with constants $V_i$ and $B_i$ which are independent of $N_{i,L}$, but depend on the value of $\mathbf{x}_i$. For the second inequalities, we assumed that both bias and variance are bound for all $\mathbf{x}$ and the corresponding constants $V$ and $B$ are simply the maximum of the respective pointwise bound constants and therefore independent of the value of $\mathbf{x}_i$.

Using Equation \eqref{eq:ApproximateFunctionValues}, we can split the interpolation ${\tilde f}_L$ on basis of the approximate function evaluations into three contributions
\begin{align}
 {\tilde f}_L &:= \sum\limits_{i=1}^{M_L} {\tilde f}_{i,L} \psi_{i,L} = f_L + f_{b, L} + f_{\eta, L}\\
 & \text{with} \nonumber \\
 f_{b, L}&=\sum\limits_{i=1}^{M_L} {b}_{i,L} \psi_{i,L} \label{eq:BiasInterpolation}\\
f_{\eta, L}&=\sum\limits_{i=1}^{M_L} {\eta}_{i,L} \psi_{i,L} \label{eq:NoiseInterpolation}\\
\end{align}
where $f_L$ is the interpolant \eqref{eq:SGInterpolantLagrange} of the function we want to approximate and $f_{b, L}$ and $f_{\eta, L}$ are the deviations from this due to bias and noise, respectively. The deviation from our target function $f$ thus consists of $f_{b, L}$, $f_{\eta, L}$ and $\Delta f_{L} := f-f_L$. The latter originates from the numerical error by the SG interpolation, and the former two result from the approximate Monte Carlo sampling. We measure the error $e_L$ by the root of the expected squared 2-norm difference between ${\tilde f}_L$ and $f$, i.e.
\begin{equation}
 e_L:= E(|| f - {\tilde f}_L||_2^2)^{1/2}  \label{eq:ErrorNorm}
\end{equation}
where again the expectation is with respect to the randomness of the samples. Note, that $ E(|| \cdot ||_2^2)^{1/2}$ fulfills the norm axioms.

\subsection{Error splitting for affine transformations}\label{app:error-splitting}

We first consider the case when $T$ is an affine transformation. Since the expectation is a linear map, the bias vanishes and the error $e_L$ obeys
\begin{align}
    e_L^2&= E(|| f  - {\tilde f}_L||_2^2)\label{eq:ErrorSplittingAffineElegantZero}\\
                &= E(|| \Delta f_L + f_{\eta, L}||_2^2)\label{eq:ErrorSplittingAffineNormOfSum}\\
                &= || \Delta f_L ||_2^2 + 2 E( \langle \Delta f_L, f_{\eta, L}\rangle) + E(|| f_{\eta, L}||_2^2)\label{eq:ErrorSplittingAffineScalarProduct}\\
                &= || \Delta f_L ||_2^2 + E(|| f_{\eta, L}||_2^2)\label{eq:ErrorSplittingAffineZeroCrossTerms}\\
                &= || \Delta f_L ||_2^2 + \sum\limits_{i=1}^{M_L} \text{Var}({\tilde f}_{i,L}) || \psi_{i,L}||^2_2\label{eq:ErrorSplittingAffinePointwiseVariances},
\end{align}
where $\Delta f_L=f-f_L$. Equation \eqref{eq:ErrorSplittingAffineScalarProduct} results from the definition of the norm $|| \cdot ||_2$. Equation \eqref{eq:ErrorSplittingAffineZeroCrossTerms} results from $E( \langle \Delta f_L, f_{\eta, L}\rangle)=0$ as
\begin{equation}
\begin{split}
 E( \langle \Delta f_L, f_{\eta, L}\rangle) & = E( \langle \Delta f_L, \sum\limits_{i=1}^{M_L} \eta_{i,L} \psi_{i,L}\rangle)\\
 &=\sum\limits_{i=1}^{M_L} \langle \Delta f_L,\psi_{i,L} \rangle  E(\eta_{i,L})=0
\end{split}
\end{equation}
where the second equality arises because both $ \Delta f_L$ and $\psi_{i,L}$ are independent of the samples and $E(\eta_{i,L})=0$. The identity Equation \eqref{eq:ErrorSplittingAffinePointwiseVariances} results from
 \begin{equation}\label{eq:ExactNoiseErrorAffine}
\begin{split}
  E(|| f_{\eta, L}||_2^2)&= E(\langle f_{\eta, L}, f_{\eta, L} \rangle )\\
                         &= \sum\limits_{i,j=1}^{M_L} E(\eta_{i,L} \eta_{j,L})\langle \psi_{i,L}, \psi_{j,L} \rangle )\\
                         &=  \sum\limits_{i=1}^{M_L} \text{Var}({\tilde f}_{i,L}) || \psi_{i,L}||^2_2
 \end{split}
\end{equation}
where we exploited that $\eta_{i,L}$ and $\eta_{j,L}$ have zero mean and are statistically independent for $i\neq j$ as they are calculated from independent samples. Hence, $E(\eta_{i,L} \eta_{j,L})= \text{Var}(\eta_{i,L} ) \delta_{ij}$, which, together ${\tilde f}_{i,L} = E ({\tilde f}_{i,L}) + \eta_{i,L}$, leads to Equation \eqref{eq:ExactNoiseErrorAffine}.

\subsection{Minimization of the sampling cost under error bound constraints for affine transformations}\label{app:minimisation-problem}

We assume that the cost per sample is independent of $\mathbf{x}_i$. Then, the overall cost  is proportional to the total number of sampling steps $\sum_{i=1}^{M_L} N_{i,L}$. To address the question of how to choose ${N_{i,L}}$, we consider the problem of minimizing the cost for sampling under the constraint that the noise error $E(|| f_{\eta, L}||_2^2)^{1/2}$ stays below a predefined threshold $\epsilon_L$. Using the variance bound \eqref{eq:VarianceBound}, we arrive at a bound for the noise error
\begin{align}
 E(|| f_{\eta, L}||_2^2) & \leq  V\sum\limits_{i=1}^{M_L} \frac{1}{{N}_{i,L}} || \psi_{i,L}||^2_2
\end{align}
and require this to be below the threshold. We arrive at the constraint minimization problem
\begin{align}\label{eq:minimisation-problem}
 \min\limits_{\{ {N}_{i,L}  \}} \sum\limits_{j=1}^{M_L} {N}_{j,L}\\
 \text{such that}\\
  V\sum\limits_{i=1}^{M_L} \frac{1}{{N}_{i,L}} || \psi_{i,L}||^2_2 \leq \epsilon_L^2 \\
  {N}_{i,L} >  0
\end{align}
For simplicity, we assume that ${N}_{i,L}$ are real numbers, that are then later rounded up to integers. Using the method of Lagrange multipliers, $N_{i,L}$ is the smallest integer for which holds
\begin{align}
N_{i,L}&\geq\frac{V}{\epsilon_L^2} Z_{i,L}\label{eq:ErrorMinimizationChoiceNOfSamples}\\
&\text{or}\nonumber \\
\text{Var}(\tilde{f}_{i,L}) &\leq \frac{\epsilon_L^2}{Z_{i,L}}. \label{eq:ErrorMinimizationChoiceVariance}
\end{align}
where $Z_{i,L}:=  || \psi_{i,L}||_2\sum_n  || \psi_{i,L}||_2$ which can be calculated using the definition \eqref{eq:LagrangeBasis} and the analytically accessible scalar products between the SG basis functions. Note that, Equation \eqref{eq:ErrorMinimizationChoiceVariance} is the more practical than Equation \eqref{eq:ErrorMinimizationChoiceNOfSamples} to choose $N_{i,L}$. The constants $V_i$ in the variance bound \eqref{eq:VarianceBound} can be estimated from a few samples individually for each grid point and employed to determine $N_{i,L}$. In contrast, the constant $V$ is difficult to determine without sampling the domain $\Omega$, which would contradict the purpose of on-the-fly interpolation.

By Equation \eqref{eq:SG_error_bound}, $|| \Delta f_L ||_2$ converges to zero for increasing $L$.
Further, the above arguments show, that any choice of $\epsilon_L$ together with \eqref{eq:ErrorMinimizationChoiceNOfSamples} for $N_{i,L}$ will lead to decreasing $E(|| f_{\eta, L}||_2^2)$ with $L$ as long as $\epsilon_L$ decays with $L$  (irrespective of the transform $T$ as long the delta theorem applies).
For affine transformations, the identity \eqref{eq:ErrorSplittingAffineZeroCrossTerms} then implies that the total approximation error $||f - \tilde{f}_L||$ approaches zero for $L\rightarrow \infty$ and the approach is convergent.
In practice, the choice of $\epsilon_L$ influences the efficiency of the approach. In the main text, we argued that choosing
\begin{equation}
 \epsilon_L=C_\epsilon 2^{-L}
\end{equation}
is a good compromise for balancing discretization and noise error and provided some simple heuristics to obtain reasonable choices for $C_\epsilon$. This choice leads to
\begin{equation}
 N_{i,L}\geq\frac{V}{C_\epsilon} 2^{2L} Z_{i,L}\label{eq:ErrorMinimizationChoiceNOfSamplesLinearThreshold}
\end{equation}

\subsection{Bias contribution bounds}\label{app:noise-bias}
If the stochastic model evaluations contain a bias due to application of  a general, nonlinear transform $T$, the bias contribution $f_{b,L}$ does not automatically vanish. To investigate its influence, we investigate
\begin{align}\label{eq:bias-error-splitting}
E(\|f-\tilde{f_L}\|^2_2)^{1/2} & = E(\|\Delta f_L + f_{b,L} + f_{\eta,L}\|^2_2)^{1/2}\\
& \leq \|\Delta f_L\|_2 + E(\| f_{\eta,L}\|^2_2)^{1/2} + \|f_{B,L}\|_2,
\end{align}
where we exploited the triangle inequality (since $E(\|\cdot\|^2_2)^{1/2}$ is a norm) and that $\Delta f_L$ and $f_{\eta,L}$ are not stochastic and we can drop the expectation for the corresponding terms.

Choosing $N_{i,L}$ according to the arguments in App. \ref{app:minimisation-problem}, $ \|\Delta f_L\|_2 + E(\| f_{\eta,L}\|^2_2)^{1/2}$ will approach zero for large enough $L$. The bias contribution $\|f_{B,L}\|_2$ can be bound by
\begin{align}\label{eq:bias-term-nodal-basis-bound-general}
    \|f_{B,L}\|_2&=\left\|\sum_{i=1}^{M_L} b_i \psi_{i,L} \right\|_2 \leq \sum_{i=1}^{M_L} |b_{i}| \cdot \|\psi_{i,L}\|_2 \nonumber\\
    &\leq B \sum_{i=1}^{M_L} N^{-1}_{i,L}\|\psi_{i,L}\|_2
\end{align}
where we utilized the triangle inequality and the bound \eqref{eq:BiasBound} for the bias. Choosing $N_{i,L}$ according to Eq. \eqref{eq:ErrorMinimizationChoiceNOfSamplesLinearThreshold} leads to
\begin{align}
    \|f_{B,L}\|_2& \leq \frac{B\epsilon}{V} 2^{-2L}\sum_{i=1}^{M_L} \left(\sum_{k=1}^{M_L}\|\psi_{k,L}\|_2\right)^{-1} \\
                 &= \frac{B\epsilon}{V} 2^{-2L} M_L\left(\sum_{k=1}^{M_L} \|\psi_{k,L}\|_2\right)^{-1} \\
                 & \leq O\left( 2^{-L}L^{d-1} \right)
\end{align}
where we used that $M_L =O(2^L L^{d-1})$ \cite{bun04}, and that $\left(\sum_{k=1}^{M_L} \|\psi_{k,L}\|_2\right)^{-1}\leq 1$ (cf. Equation \eqref{eq:LowerBoundNormLagrangeBasis}). The above bound converges slower than the bound for the noise error \eqref{eq:ErrorMinimizationChoiceNOfSamplesLinearThreshold}. However, Equation \eqref{eq:LowerBoundNormLagrangeBasis} is a very conservative bound. We thus expect that the bias bound generally converges faster than $O\left( 2^{-L}L^{d-1} \right)$.

\section{Reverse water gas shift on CuZnO} \label{app::WGS_CuZnO}
The reactivity for the reverse water gas shift is modeled using a literature \cite{bussche1996steady} analytical expression for the reactivity
\begin{align}\label{eq:rwgs-R}
    R_{\text{RWGS}} = \frac{k p_{\ch{CO2}} \cdot \left(1- \frac{p_{\ch{CO}} \cdot p_{\ch{H2O}}}{K_{5}\cdot p_{\ch{CO2}} \cdot p_{\ch{H2}}} \right)}
    {1 + K_2 \cdot \frac{p_{\ch{H2O}}}{p_{\ch{H2}}}+ K_3 \cdot (p_{\ch{H2}})^{0.5} +K_4 \cdot p_{\ch{H2O}}}
\end{align}
where $\nu_\alpha$ is the stoichiometric coefficient, and \( p_{\text{CO}} \),  \( p_{\text{CO}_2} \),$p_{\ch{H2O}}$, $p_{\ch{H2}}$ are the partial pressures of \ch{CO}, \ch{CO2}, \ch{H2O} and \ch{H2} respectively.\\
The rate constant $k$ and the equilibrium constants $K_{i}$ are given by
\begin{align}\label{eq:rwgs-constants}
k &= A \cdot e^{-\frac{E_{a,i}}{R_{\text{gas}}T}} \nonumber \\
K_{i} &= k_i \cdot e^{\frac{E_{r,i}}{R_{\text{gas}} T}}
\end{align}
with constants $A,E_{a/r,i},k_i$ corresponding to \cite{brosigke2022closer}, where  $R_{\text{gas}}$ denotes the universal gas constant and $T$ the temperature. 

\section{CO-oxidation model on RuO$_2$} \label{app:CO-RuO2}
As a CO oxidation model, we take a well known model from the literature \cite{reuter2004steady}.
The model describes CO-oxidation on the RuO2(110) surface. We simulate a 20x20 lattice grid with two types of active sites. For each evaluation, we employ 1E6 kmc steps for relaxation and $10^6$ steps for sampling at each evaluation of the model.
Each site of the grid can be either occupied by \ch{CO} or \ch{O} or be empty. In total, the model consists of 22 possible processes, including non-dissociative CO adsorption/desorption, dissociative O2 adsorption/desorption, diffusion of \ch{CO} and \ch{O} and \ch{CO2} formation.
The mechanisms, corresponding energies and more in depth discussions can be found in the literature, e.g. \cite{reuter2001composition}. 
We used the implementation of the model in kmos3 \cite{hoffmann2014kmos}.

\section{Water-gas-shift model} \label{app::WGS}
We employ a model for the water gas shift reaction on \ch{Pt}(111). The model is adapted from the 1p-kMC model  water gas shift model from reference \cite{stamatakis2011graph}, where we omitted lateral interaction for simplicity.
This model lives on a hexagonal lattice with one type of site per surface  unit cell.
The rate constant $k$ of an elementary process has Arrhenius form, i.e., $k=Ae^{-\frac{E_a}{k_b T}} p_\alpha$ for a process with the gaseous species $\alpha$ among the reactants where $p_\alpha$ is the corresponding partial pressure. For a process without gaseous species among the reactants, the rate constant i simply $k=Ae^{-\frac{E_a}{k_b T}}$ for usual processes. The elementary processes and corresponding activation energies $E_a$ and prefactors $A$ can be found in Table \ref{tab:mechanism-WGS}.

In the simulation, we employ $20 \times 20$ grid of equivalent surface unit cells, that can be occupied by one of the surface species or be empty, and $2\times 10^6$ kMC steps for relaxation and $2\times 10^6$ steps for sampling at each evaluation of the model.
\onecolumn
\begin{table}
\centering
    \begin{tabular}{|l|l|l|l|}
\hline
 process name &  & $E_a[\si{eV}]$ & A \\
\hline
CO Adsorption    &\ch{CO + ^*  \, -> \, ^*CO} &  0.00 & 6.816 $\times 10^{5} \si{bar^{-1} s^{-1}}$\\
CO Desorption    &\ch{^*CO   \, -> \, CO + ^*} & 1.17 & 1.989 $\times 10^{14} \si{ s^{-1}}$ \\
\ch{H2O} Adsorption    &\ch{H2O + ^*  \,-> \, ^*H2O} &  0.00&7.204 $\times 10^{1} \si{bar^{-1} s^{-1}}$\\
\ch{H2O} Desorption    &\ch{^*HO \,-> \,  H2O + ^* } &  0.29&4.260 $\times 10^{8} \si{ s^{-1}}$\\
\ch{O2} Adsorption    &\ch{O2 + ^* + ^*  \,-> \, ^*O + ^*O} &  0.00& 2.356 $\times 10^{8}\si{bar^{-1} s^{-1}}$\\
\ch{O2} Desorption   &\ch{^*O + ^*O \,-> \, O2 + ^* + ^*} &  2.32&4.864 $\times 10^{15} \si{ s^{-1}}$\\
\ch{H2} Adsorption  &\ch{H2 + ^*  + ^*  \,-> \, ^*H  + ^*H} &  0.00&5.746 $\times 10^{7} \si{bar^{-1} s^{-1}}$\\
\ch{H2} Desorption  &\ch{^*H + ^*H  \,-> \, H2  + ^* + ^*} &  1.00&1.235 $\times 10^{13} \si{ s^{-1}}$\\
\ch{CO2} Adsorption    &\ch{CO2 + ^* + ^*  \,-> \, ^*CO + ^*O} &  1.23& 6.654 $\times 10^{4} \si{bar^{-1} s^{-1}}$\\
\ch{CO2} Desorption   &\ch{^*CO + ^*O  \,-> \, CO2  + ^*} &  1.11&1.175 $\times 10^{12} \si{bar^{-1} s^{-1}}$\\
\ch{H2O} decomposition   &\ch{^*H2O + ^*  \, -> \, ^*OH + ^*H} & 0.65 &4.475 $\times 10^{11}\si{s^{-1}}$ \\
\ch{H2O} composition   & \ch{^*OH + ^*H \, -> \, ^*H2O + ^*} &   0.25&7.001 $\times 10^{10} \si{s^{-1}}$\\
\ch{OH} decomposition   &\ch{^*OH + ^*  \,-> \, ^*O + ^*H} &  0.79&2.360 $\times 10^{13} \si{s^{-1}}$\\
\ch{OH} composition   & \ch{^*O + ^*H \, -> \, ^*OH + ^*} &   1.10&1.194 $\times 10^{12} \si{s^{-1}}$\\
\ch{OH,OH} disproportionation   & \ch{^*OH + ^*OH \, -> \, ^*H2O + ^*O} &   0.00&3.088 $\times 10^{11} \si{s^{-1}}$\\
\ch{OH,OH} comproportionation   & \ch{^*H2O + ^*O \, -> \, ^*OH + ^*OH} &  0.71 & 9.984 $\times 10^{10} \si{s^{-1}}$\\
\ch{COOH} decomposition   &\ch{^*COOH + ^*  \,-> \,CO2 + ^*H +  ^* } &  0.67&5.280 $\times 10^{14} \si{s^{-1}}$\\
\ch{COOH} composition   & \ch{CO2 + ^* + ^*H  \, -> \, ^*COOH + ^*} &    0.63& 5.895 $\times 10^{4} \si{bar^{-1} s^{-1}}$\\
\ch{CO,OH} disproportionation   & \ch{^*CO + ^*OH \, -> \, ^*COOH + ^*} & 0.48& 4.582 $\times 10^{11} \si{s^{-1}}$ \\
\ch{CO,OH} comproportionation  & \ch{^*COOH + ^* \, -> \, ^*CO + ^*OH} & 0.95 &1.175 $\times 10^{13} \si{s^{-1}}$\\
\hline
\end{tabular}
\caption{Mechanisms, Energies and prefactors for the water-gas-shift model}
\label{tab:mechanism-WGS}
\end{table}

\medskip

\printbibliography
\end{document}